\documentclass[review,3p,times]{elsarticle}

\usepackage{amssymb}
\usepackage{textcomp, gensymb}
\usepackage{upgreek}
\usepackage{multirow}

\usepackage{lineno}

\usepackage{color,soul}

\journal{Materialia}

\begin{document}

\begin{frontmatter}



\title{High-throughput assessment of the microstructural stability of segregation-engineered nanocrystalline Al-Ni-Y alloys}


\author[UCSB:Mat]{W. Streit Cunningham}
\author[UCSB:Mat,GWNU]{Jungho Shin}
\author[UCI:Mat]{Tianjiao Lei}
\author[UCI:Mat]{Timothy J Rupert}
\author[UCSB:Mat]{Daniel S. Gianola\corref{cor1}}

\affiliation[UCSB:Mat]{organization={Materials Department},
            addressline={University of California}, 
            city={Santa Barbara},
            postcode={93106}, 
            state={CA},
            country={USA}}
\affiliation[GWNU]{organization={Gangneung-Wonju National University},
            city={Gangneung-si},
            state={Gangwon-do},
            country={Republic of Korea}}
\affiliation[UCI:Mat]{organization={Department of Materials Science and Engineering},
            addressline={University of California}, 
            city={Irvine},
            postcode={92697}, 
            state={CA},
            country={USA}}

\cortext[cor1]{Corresponding author}

\begin{abstract}
Segregation engineering has emerged as a promising pathway towards designing thermally stable nanocrystalline alloys with enhanced mechanical properties.
However, the compositional and processing space for solute stabilized microstructures is vast, thus the application of high-throughput techniques to accelerate optimal material development is increasingly attractive. 
In this work, combinatorial synthesis is combined with high-throughput characterization techniques to explore microstructural transitions through annealing of a nanocrystalline ternary Al-base alloy containing a transition metal (TM=Ni) and rare earth dopant (RE=Y).
A down-selected optimal composition with the highest thermal stability is annealed through \textit{in situ} transmission electron microscopy, revealing that the removal of the RE species is correlated to a reduction in the microstructural stability at high temperatures as a result of variations in intermetallic phase formation.
Results demonstrate the benefits of co-segregation for enhancing mechanical hardness and delaying the onset of microstructural instability. 
\end{abstract}

\begin{graphicalabstract}
\includegraphics{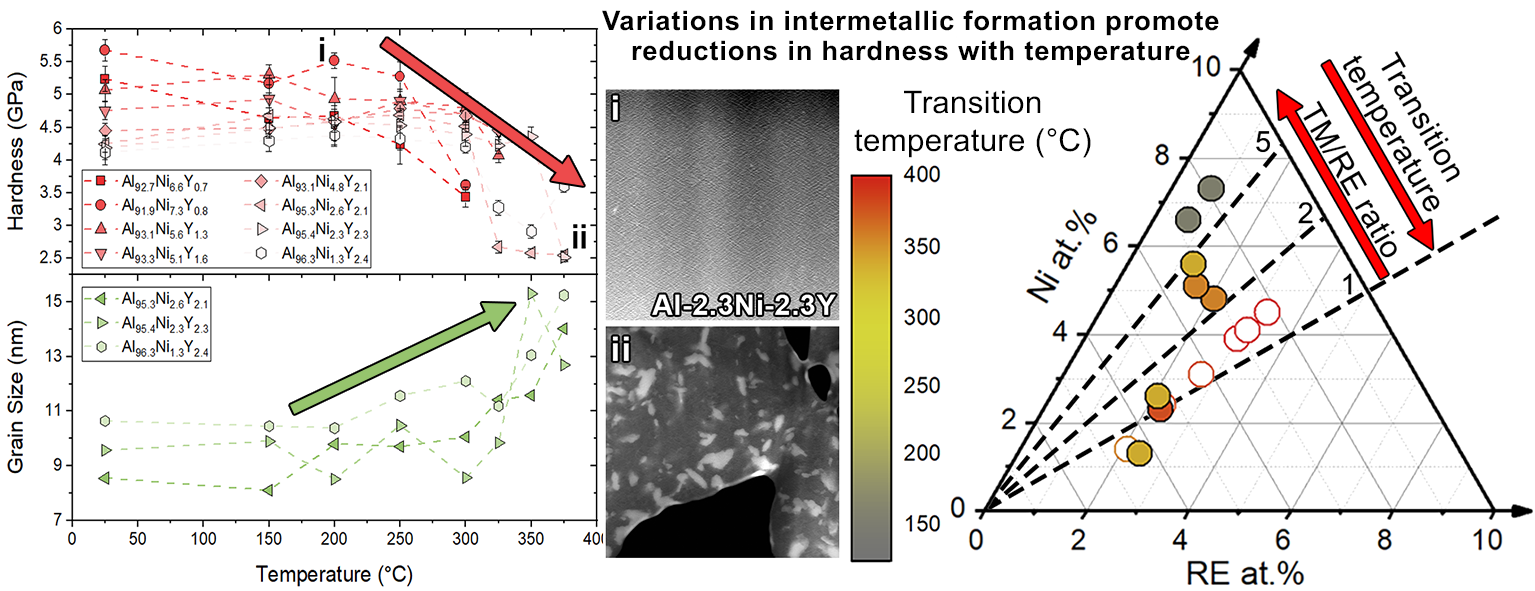}
\end{graphicalabstract}



\end{frontmatter}

A significant body of literature has focused on the development of nanocrystalline metals for their enhanced hardness \cite{KUMAR:2003,Hahn:2015}, improved wear resistance \cite{Padilla:2010,Rupert:2013}, and promising irradiation behavior \cite{Beyerlein:2013} relative to traditional coarse-grained metals. 
However, the poor microstructural stability arising from their far-from-equilibrium state threatens broad applicability of nanocrystalline metals, as the high excess Gibbs free energy at the grain boundary frequently drives significant grain coarsening even at low thermal loads \cite{ANDRIEVSKI:2014}. 
Segregation engineering of nanocrystalline microstructures remains a promising pathway towards stabilizing nanocrystalline metals through reduction of the driving force for grain growth via manipulation of the thermodynamic state of the system \cite{Chookajorn:2012,Darling:2013,Detor:2007} and/or through restriction of the grain boundary kinetics \cite{El-Sherik:1991,Morris:1991}. 
While compelling progress has been made on solute stabilization of binary nanocrystalline alloys \cite{Cunningham:2022,Amram:2018,Chookajorn:2014,Khalajhedayati:2015,Koch:2008}, recent work on increasingly complex higher component systems has yielded even higher efficacies relative to elemental or binary nanocrystalline alloys \cite{XING:2018,KUBE:2020,LEI:2021}.
Through increasing microstructural complexity, a notable pathway towards even better microstructural stability that utilizes unique grain boundary states such as amorphous intergranular films, also termed amorphous complexions, has been shown in recent works to result in exceptional thermal stability and enhanced mechanical behavior \cite{Khalajhedayati:2015,cantwell2014grain, cantwell2020grain,Balbus:2021,Shin:2022}.
\par

While increasing the number of alloying elements has yielded promising improvements in microstructural stability, exploring the vast compositional and processing space for achieving optimal segregation-engineered nanocrystalline alloys is a significant undertaking. 
To that end, higher throughput techniques for mapping structure-property relationships, in combination with combinatorial synthesis, have arisen as a promising approach for rapid and accurate determination of material properties through composition space.
Indeed, successful applications of combinatorial approaches have been demonstrated across a wide range of systems, including multi-principal element alloys (MPEAs) \cite{Li:2018}, metallic glasses \cite{Ding:2012}, and nanocrystalline alloys \cite{KUBE:2020,Shin:2022, li2021achieving}.
Recent work on amorphous complexion-containing ternary Al alloys with transition metal (TM) and rare earth (RE) dopants, Al-TM-RE, have exhibited exceptional high-temperature stability \cite{Balbus:2021} and improved mechanical properties \cite{Balbus:2020,Shin:2022}, with increasing dopant concentrations yielding consistently improved behaviors. 
However, while increasing concentrations of RE dopants typically results in enhanced material properties, RE elements are prohibitively expensive, limiting their applicability in real spaces.  
Other materials sustainability factors such as those captured by the Herfindahl-Hirschman Index (HHI) point to market concentration and scarcity risks, particularly for several of the attractive alloying elements for lightweight structural materials that impart beneficial properties \cite{gaultois2013data, mansouri2017balancing}, thereby motivating the need for careful compositional optimization. 
In this work, a combinatorial approach is applied to a ternary nanocrystalline Al-TM-RE system to investigate the influence of reducing the RE content on the overall mechanical properties.
An examplar nanocrystalline Al-base alloy, Al-Ni-Y, was selected for its strong amorphous complexion-forming behavior, its manufacturability in bulk form \cite{LEI:2021}, and promising mechanical property retention when subjected to thermal exposures.  Y, in particular, appears to be an excellent alloying addition based on its strong stabilizing effect and amorphous complexion formation, but its market concentration is the highest of all elements according to the HHI index \cite{mansouri2017balancing}; we seek to optimize the thermomechanical properties while minimizing the total amount of alloying additions.

Nanocrystalline Al-Ni-Y alloys were synthesized on (100) Si wafers using an AJA ATC 1800 sputter deposition system; details of the deposition can be found in Ref. \cite{Shin:2022}.
Indentation tests were performed using a KLA iMicro Nanoindenter equipped with a 50 mN load cell and a diamond Berkovich tip using the procedure outlined in Ref. \cite{Shin:2022}.
XRD scans were performed on a Rigaku SmartLab diffractometer using a Cu K$\alpha$$_1$ source. 
Lattice parameter (\emph{a}) and mean grain size (\emph{d}) of the Al-rich alloys were determined from fitting the profile of the \{111\} peak \cite{Cheary:hw0001} and the coherently scattered length from the Scherrer equation \cite{Patterson:1939}.  Residual stresses were found to be negligible based on previous work on similar systems \cite{Shin:2022}.
Samples for transmission electron microscopy (TEM) were prepared through a focused ion beam (FIB) lift-out procedure in an FEI Helios Dualbeam Nanolab 600.
STEM high angle annular dark field (STEM-HAADF) and STEM dark field (STEM-DF) micrographs, and complementary energy-dispersive X-ray spectroscopy (STEM-EDS) maps, were collected using a ThermoFisher Talos G2 200X TEM/STEM operated at 200 keV with collection angles for STEM-HAADF and STEM-DF of 61–200 and 12–20 mrad, respectively.
\textit{In situ} annealing was performed in the ThermoFisher Talos using a MEMS-based Protochips Fusion holder.
The sample was annealed to consecutively higher temperatures (175, 200, 250, 300 \degree C) using a heating rate of 100 \degree C/s, followed by an isothermal hold of 5 mins and immediate quenching at 500 \degree C/s for acquisition of STEM-EDS maps.
Videos were acquired at a 1024x1024 pixel resolution and with dwell times of 500 ns, resulting in a total frame time of 780 ms, or roughly 1 fps. 
\par

\begin{figure}[ht] 
    \centering
    \includegraphics[width=0.5\textwidth]{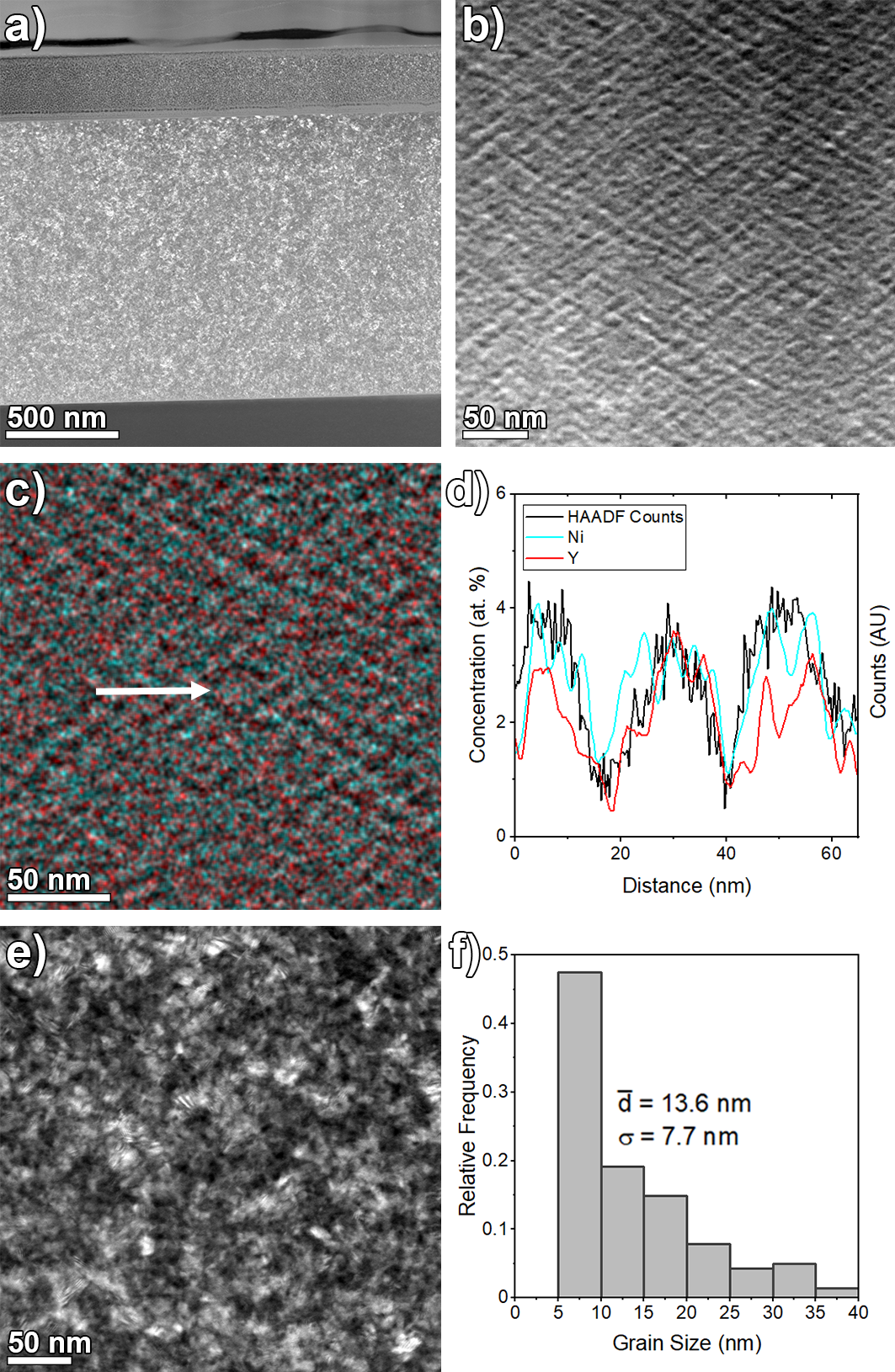}
    \caption{(a) STEM-DF micrograph of Al-2.3Ni-2.3Y annealed at 200 \degree C. (b) STEM-HAADF micrograph and corresponding (c) STEM-EDS map of Ni (light blue) and Y (red) for Al-2.3Ni-2.3Y annealed at 200 \degree C. A line-scan (d) across three grain boundaries (white arrow in (c)) indicates grain boundary segregation of the two solutes. (e) Higher magnification STEM-DF micrograph emphasizing nanocrystallinity of the microstructure. (f) Distribution of grain sizes for the microstructure shown in (e).}
    \label{fig:fig1}
\end{figure}

Stabilization of nanograins and grain boundary enrichment is evident, as demonstrated in a STEM-DF micrograph of a representative Al-TM-RE alloy, Al-2.3 at.\%Ni-2.3 at.\%Y, annealed at 200 \degree C (\textbf{Figure \ref{fig:fig1}a}). Inspection of the microstructure confirms a sputtered film thickness of 1 $\upmu$m and grain sizes below 100 nm. 
Variations in the contrast shown through STEM-HAADF in \textbf{Figure \ref{fig:fig1}b} suggest compositional variations across the microstructure, which are verified by a line-scan (\textbf{Figure \ref{fig:fig1}d}) across the corresponding STEM-EDS map (\textbf{Figure \ref{fig:fig1}c}), confirming co-segregation of the dopant atoms, Ni and Y, at the grain boundaries. 
Nanocrystalline grain sizes of 13.6$\pm$7.7 nm were measured through analysis of the STEM-DF micrograph shown in \textbf{Figure \ref{fig:fig1}e}, with the largest grains approaching $\sim$40 nm (grain size distribution provided in \textbf{Figure \ref{fig:fig1}f}). 
Similarities between the average grain size and wavelength of the dopant distribution along the STEM-EDS line-scan further indicate chemical segregation at the grain boundaries.
Such trends are consistent with prior work on Al-Ni-Ce, where relaxation of the microstructure during low temperature annealing was shown to drive dopant segregation at the grain boundaries \cite{Shin:2022}. 
\par

\begin{figure}[ht] 
    \centering
    \includegraphics[width=0.9\textwidth]{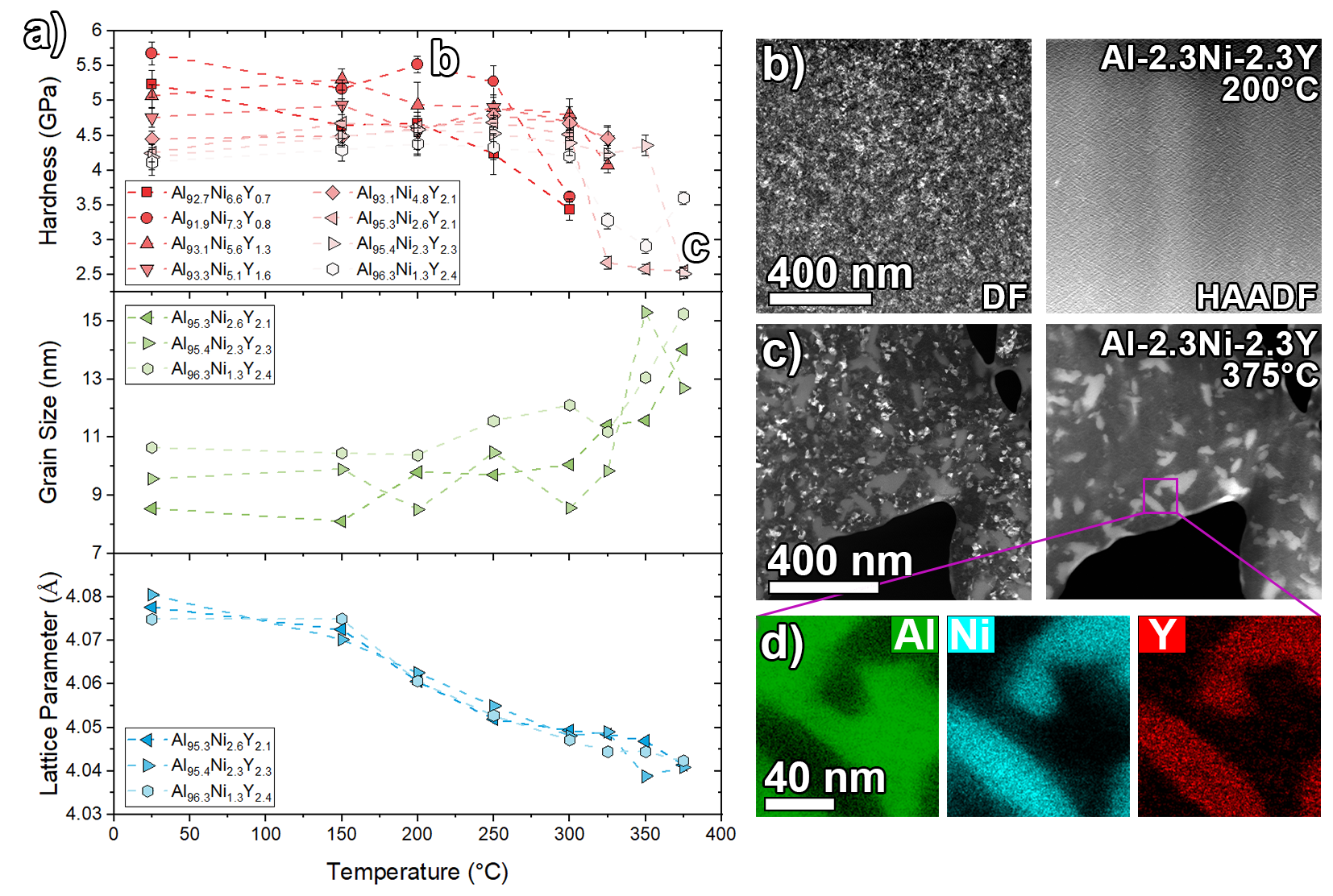}
    \caption{(a) Hardness as a function of temperature for all Al-Ni-Y samples in this study and corresponding grain sizes/lattice parameter for three representative Al-Ni-Y microstructures (Al-2.6Ni-2.1Y, Al-2.3Ni-2.3Y, and Al-1.3Ni-2.4Y). STEM-DF and STEM-HAADF micrographs of Al-2.3Ni-2.3Y after annealing at (b) 200 \degree C and (c) 375 \degree C. (d) STEM-EDS maps (Al, Ni, and Y) of intermetallics in Al-2.3Ni-2.3Y annealed at 375 \degree C.}
    \label{fig:fig2}
\end{figure}

To determine the optimal combination of alloying elements, eight different compositions of Al-Ni-Y ranging from Ni + Y = 3.7 at.\% to Ni + Y = 8.1 at.\% were annealed \textit{ex situ} to various temperatures up to 400 \degree C (72\% of the melting point of Al); all of the Al-Ni-Y compositions and annealing temperatures are laid out in \textbf{Table \ref{tab:tab1}}.
Mechanical hardness was probed through nanoindentation after each annealing step to capture mechanical property changes, with the results for each composition provided in \textbf{Figure \ref{fig:fig2}a}. 
All samples maintained an average hardness of 4-6 GPa below annealing temperatures of 250 \degree C, with increasing combined dopant concentrations (Ni + Y) generally resulting in a higher average hardness.
This hardness range is several times higher than pure nanocrystalline Al \cite{Pun:2017} and far exceeds the hardness of most commercial Al alloys \cite{Davis:1993}.
To confirm the nanocrystallinity of the microstructures, average grain size and lattice parameter for the three lowest combined dopant concentrations (Ni + Y = 4.7, 4.6, and 3.7 at.\%) were estimated through fitting of the FCC Al peak from XRD and are plotted in \textbf{Figure \ref{fig:fig2}a}.
XRD results confirm nanocrystalline grain sizes below 11 nm for all three compositions until 250 \degree C, correlating closely with the trends observed in the hardness.
On the other hand, lattice parameters for the same compositions decrease slowly from approximately 4.08 \AA{} at room temperature to 4.075 \AA{} at 150 \degree C, and decrease rapidly to 4.055 \AA{} at 250 \degree C.
Noting that the lattice parameter of pure FCC Al is 4.050 \AA{}, reductions in the lattice parameter at low annealing temperatures suggests relaxation of the grain interiors through dopant co-segregation towards the grain boundaries as corroborated by the STEM-EDS maps in \textbf{Figure \ref{fig:fig1}c}.
\par

\begin{table}[ht]
    \centering
    \begin{tabular}{c|cccccccc}
    & \multicolumn{8}{c}{Composition (at.\%)}               \\ \hline
    Al      & 91.9 & 92.7 & 93.1 & 93.1 & 93.3 & 95.3 & 95.4 & 96.3 \\
    Ni      & 7.3  & 6.6  & 5.6  & 4.8  & 5.1  & 2.6  & 2.3  & 1.3 \\
    Y       & 0.8  & 0.7  & 1.3  & 2.1  & 1.6  & 2.1  & 2.3  & 2.4 \\ 
    Ni/Y    & 9.1 & 9.4 & 4.3 & 2.3 & 3.2 & 1.2 & 1.0 & 0.5 \\ \hline 
    Temperatures (\degree C) & \begin{tabular}[t]{@{}l@{}}30\\ 150\\ 200\\ 250\\ 300\end{tabular} & \begin{tabular}[t]{@{}l@{}}30\\ 150\\ 200\\ 250\\ 300\end{tabular} & \begin{tabular}[t]{@{}l@{}}30\\ 150\\ 200\\ 250\\ 300\\ 325\end{tabular} & \begin{tabular}[t]{@{}l@{}}30\\ 150\\ 200\\ 250\\ 300\\ 325\end{tabular} & \begin{tabular}[t]{@{}l@{}}30\\ 150\\ 200\\ 250\\ 300\\ 325\end{tabular} & \begin{tabular}[t]{@{}l@{}}30\\ 150\\ 200\\ 250\\ 300\\ 325\\ 350\\ 375\end{tabular} & \begin{tabular}[t]{@{}l@{}}30\\ 150\\ 200\\ 250\\ 300\\ 325\\ 350\\ 375\end{tabular} & \begin{tabular}[t]{@{}l@{}}30\\ 150\\ 200\\ 250\\ 300\\ 325\\ 350\\ 375\end{tabular}
    \end{tabular}
    \caption{Atomic compositions for the eight Al-Ni-Y ternary alloys along with their annealing temperatures.}
    \label{tab:tab1}
\end{table}

Annealing beyond 250 \degree C led to an appreciable reduction in hardness below 4 GPa across all compositions that was concomitant with an increase in the average grain size to $\sim$15 nm.
Such trends are commonly observed in nanocrystalline metals and are typically a reflection of the reduced number of grain boundary-mediated mechanisms that accompany the reduction in grain boundary volume.
However, the increase in the average grain size is subtle, increasing by fewer than 5 nm, and thus it is unlikely that such a pronounced reduction in hardness is solely related to the increase in grain size.
To examine this reduction in hardness, TEM samples were prepared for Al-2.3Ni-2.3Y annealed at two different temperatures, 200 \degree C and 375 \degree C, and STEM-DF and STEM-HAADF micrographs for each sample are provided in \textbf{Figure \ref{fig:fig2}b} and \textbf{\ref{fig:fig2}c}, respectively.
At 200 \degree C, the microstructure is clearly nanocrystalline and in strong agreement with the average grain size estimated through XRD.
At 375 \degree C however, the microstructure has evolved significantly.
Contrast variations in the STEM-HAADF micrograph suggests the presence of secondary intermetallic precipitates throughout the bulk nanocrystalline Al matrix, with some subtle increases in the grain size observable in the corresponding STEM-DF micrograph.
Furthermore, compositional maps generated through STEM-EDS, shown in \textbf{Figure \ref{fig:fig2}d}, confirm the chemical partitioning of Ni and Y to the intermetallics.
The presence of both dopant species within the intermetallics suggests a transition in the dominant diffusion pathway and local phase equilibria with temperature: at lower temperatures, the dopants co-segregate towards the grain boundaries, while at higher temperatures, these enriched regions promote intermetallic formation \cite{mantha2021grain,devaraj2019grain, zhao2018segregation}.
 \par

\begin{figure}[ht] 
    \centering
    \includegraphics[width=0.9\textwidth]{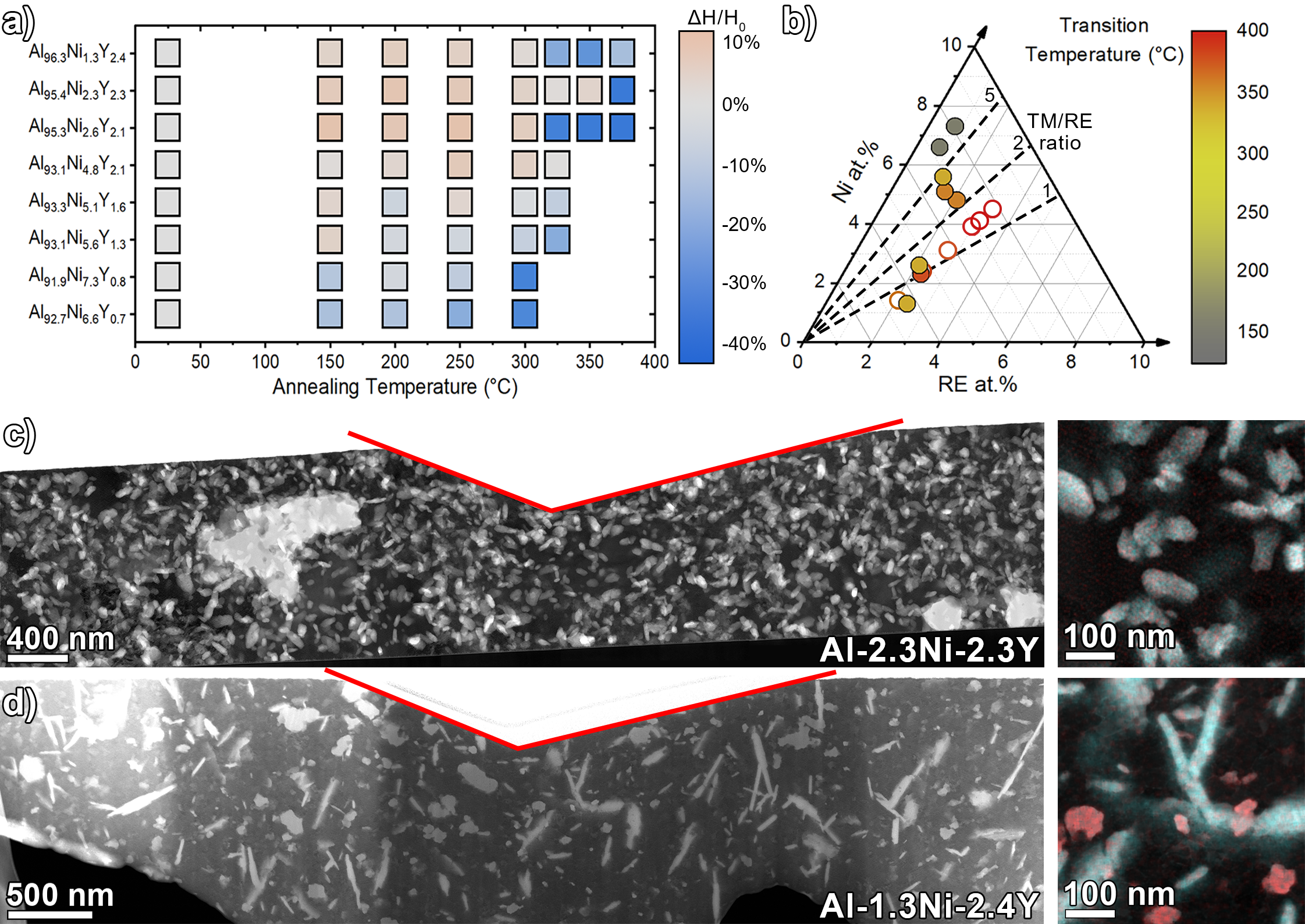}
    \caption{(a) Relative hardness vs. annealing temperatures for all Al-Ni-Y compositions. (b) Gibbs triangle of retention temperatures, or the temperature at which the hardness decreases relative to as-deposited values. Data for Al-Ni-Y represented with closed circles and reference data for Al-Ni-Ce (from Ref. \cite{Shin:2022}) represented with open circles (with the line color representing the transition temperature). Reference lines for provided for TM/RE ratios of 1, 2, and 5. STEM-HAADF micrographs of two representative Al-Ni-Y systems, Al-2.3Ni-2.3Y (c) and Al-1.3Ni-2.4Y (d), annealed at 495 \degree C with corresponding EDS maps containing Ni (in light blue) and Y (in red). TEM lamella were lifted out from within a nanoindentation indent; indent shape outlined in red.}
    \label{fig:fig3}
\end{figure}

To examine the influence of composition on the mechanical hardness and its retention following thermal exposure, the relative hardness (normalized change in hardness, $\Delta H/H_0$, where $H_0$ is the hardness of the as-deposited alloy) was plotted through annealing temperatures for all compositions of Al-Ni-Y (\textbf{Figure \ref{fig:fig3}a}). 
All compositions exhibited a reduction in the hardness of up to 40\% above some threshold temperature; the exact temperature at which each composition exhibited a drop in hardness below the initial hardness, i.e. $\Delta H/H_0 < 0$, is denoted as the "retention temperature." 
At the highest dopant concentrations, Ni+Y $>$ 7\%, and where the ratio of Ni/Y was at least 5, the hardness decreased immediately at the lowest temperature applied in this study, 150 \degree C.
Reduction of the Ni/Y ratio to between 2 to 5 led to a subtle increase in hardness at temperatures between 150 \degree C and 250 \degree C, likely related to dopant co-segregation at the grain boundaries and corresponding structural relaxation, followed by higher retention temperatures of approximately 300 \degree C relative to the highest Ni/Y ratios.
Further reduction of the Ni/Y ratio to below 2 resulted in the largest relative gains in the mechanical hardness, $\sim$10\%, at the lowest temperatures.
The highest retention temperature was observed in the case where the Ni/Y ratio approached unity.
To capture the shift in retention temperature with dopant concentration, retention temperatures were plotted as a function of Ni and Y in a Gibbs triangle shown in \textbf{Figure \ref{fig:fig3}b}.
Retention temperatures for another Al-TM-RE system, Al-Ni-Ce, published in Ref. \cite{Shin:2022}, were included to provide a comparison to an Al-TM-RE system where the TM/RE ratio was maintained at unity with increasing dopant concentrations, although other alloying ratios were not explored in that study.
As observed in Al-Ni-Ce, and at the equivalent dopant concentrations in Al-Ni-Y, equal amounts of the TM and RE species resulted in higher retention temperatures, or a delayed reduction in the hardness with increasing concentrations.
However, the removal of the RE species in Al-Ni-Y culminated in the inverse behavior, with significantly reduced retention temperatures (below 200 \degree C for Ni/Y $>$ 5) approaching three times lower than the highest retention temperature of compositions with equal amounts of TM and RE (375 \degree C for Ni/Y $=$ 1).
\par

Further examination of the microstructure after high temperature annealing was performed to explore the origins of the variation in retention temperature across compositions. 
Two compositions, Al-2.3Ni-2.3Y (Ni/Y=1) and a composition with unequal dopant concentrations, Al-1.3Ni-2.4Y (Ni/Y=0.54), were annealed to 495 \degree C (well above retention temperatures) and examined directly below the indented region to assess microstructure evolution directly at the sites corresponding to reductions in hardness, with STEM-HAADF micrographs for both compositions provided in \textbf{Figure \ref{fig:fig3}c} and \textbf{\ref{fig:fig3}d} respectively.
In Al-2.3Ni-2.3Y, the FCC Al nanocrystalline grains have significantly increased in size, with a high volume density ($\rho = 2.7\times 10^{-6} \pm 5.0\times 10^{-7}$ cm$^{-3}$) of equiaxed intermetallics throughout the sample.
Compositional analysis of the intermetallics through STEM-EDS in \textbf{Figure \ref{fig:fig3}c} show a single intermetallic composition, Al\textsubscript{9}Ni\textsubscript{3}Y (confirmed through STEM-EDS), with no segregation of Ni or Y evident elsewhere in the microstructure, indicating that the initial grain boundary enrichment gives way to partitioning to the intermetallics during precipitation.
In the sample with unequal dopant concentrations (Al-1.3Ni-2.4Y), the original nanocrystalline microstructure is also no longer evident and now consists of a bimodal distribution of intermetallics of either equiaxed or elongated shapes with a total volume density of $9.9\times 10^{-7} \pm 3.0\times 10^{-7}$ cm$^{-3}$.
STEM-EDS analysis of these intermetallics reveals two unique compositions: a binary Al\textsubscript{3}Y phase with equiaxed structures and a ternary intermetallic with elongated structures that differs in composition from the ternary intermetallic found in Al-2.3Ni-2.3Y.
Prior work on similar ultrafine-grained Ni-lean Al-Ni-Y systems have identified identical intermetallic features as embedded Al\textsubscript{23}Ni\textsubscript{6}Y\textsubscript{4} (monoclinic) in a Al\textsubscript{19}Ni\textsubscript{5}Y\textsubscript{3} (orthorhombic) matrix \cite{VASILIEV:2004}. 
Furthermore, the formation of unique intermetallic phases following minor variations in composition is consistent with experimentally determined ternary phase diagrams of Al-Ni-Y, with a two-phase equilibrium field present near compositions with TM/RE = 1 and a stable three-phase equilibrium field with TM/RE $< 1$ \cite{Mika:2010}.
\par

In the Al-Ni-Y system, variations in the mechanical hardness through annealing can thus be generally attributed to increases in grain size and the formation of intermetallics. 
However, while STEM-HAADF micrographs of the microstructures at 495 \degree C do indicate the presence of exponential grain growth (\textbf{Figure \ref{fig:fig3}c} and \textbf{\ref{fig:fig3}d}) after annealing at this very high temperature, the subtle shift in grain sizes near the retention temperatures are unlikely to be strongly correlated to the precipitous drop in hardness.
Furthermore, the range of grain sizes lies within the expected range for the inverse Hall-Petch relationship, where a shift to grain boundary mediated plasticity results in a reduction in hardness with grain size \cite{Cordero:2016}.
We note that solid solution strengthening is also unlikely to be a dominant mechanism at any composition given the significant segregation of dopants at all temperatures. 
It follows that the reduction in hardness is most strongly correlated to the onset of intermetallic formation and coarsening, as shown in \textbf{Figure \ref{fig:fig2}a}.
While it is expected for intermetallic phases to be harder than the FCC Al matrix, the uneven distribution of intermetallics could promote shear localization and subsequent softening through intermetallic-free pathways, as demonstrated in Ref. \cite{LEI:2021}.
Given the lower density of intermetallics with disparate morphologies evident in Al-Ni-Y systems with uneven dopant concentrations (\textbf{Figure \ref{fig:fig3}d}), it follows that these systems will exhibit an earlier onset of softening, as evidenced by the significant reduction in the retention temperature at the highest dopant concentrations.
\par

\begin{figure}[htbp] 
\centering
    \includegraphics[width=0.5\textwidth]{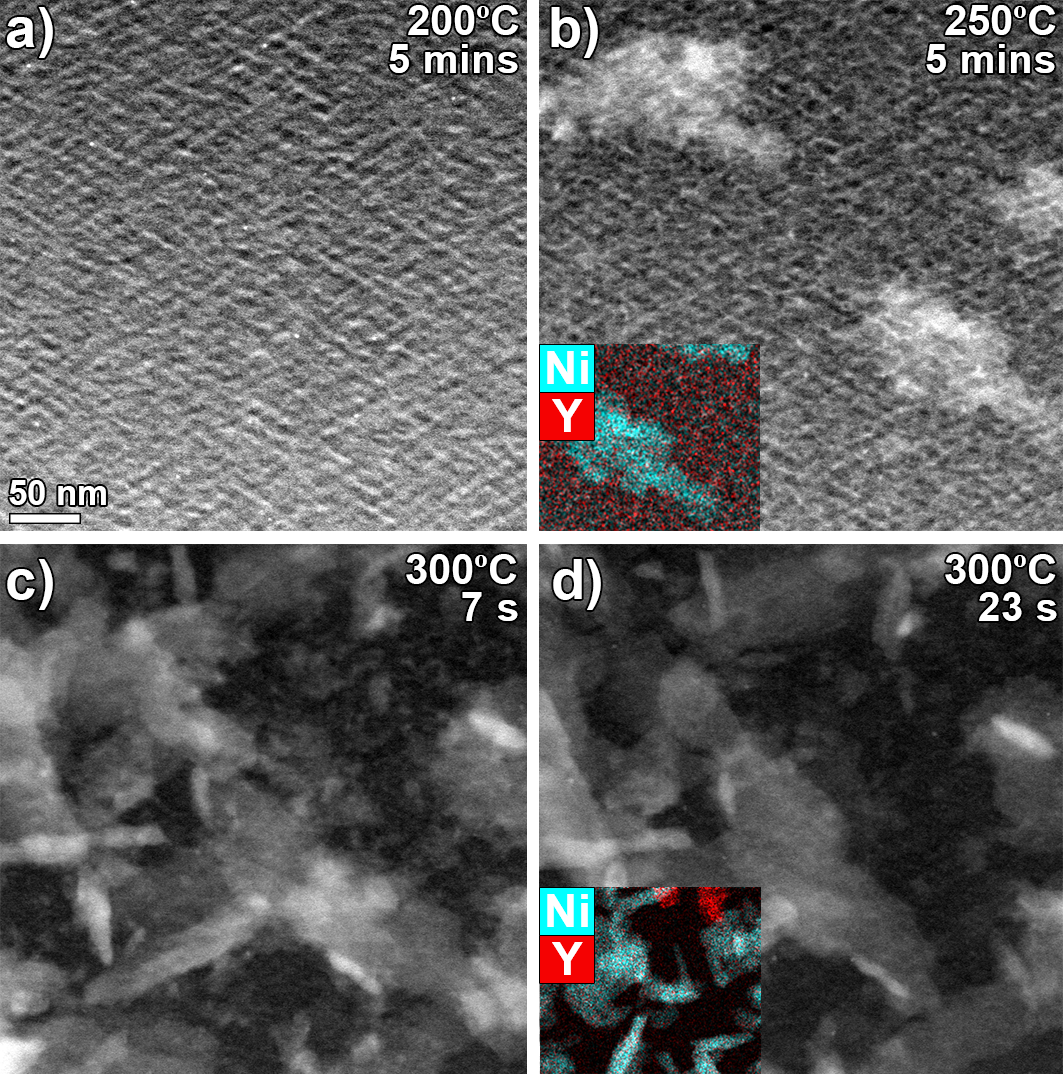}
    \caption{STEM-HAADF micrographs of Al-2.3Ni-2.3Y from \textit{in situ} heating experiment acquired (a) during annealing at 200 \degree C after 5 min, (b) during annealing at 250 \degree C after 5 min, (c) during annealing at 300 \degree C after 7s, and (d) during annealing at 300 \degree C after 23s. Representative STEM-EDS maps of Ni and Y post-quenching are provided in (b) and (d).}
    \label{fig:fig4}
\end{figure}

To directly observe the onset of intermetallic formation and its correlation to the retention temperature associated with microstructural and mechanical evolution, a TEM lamella of Al-2.3Ni-2.3Y was prepared and mounted for \textit{in situ} annealing up to 300 \degree C.
No appreciable microstructural changes were observed up to 200 \degree C (STEM-HAADF micrograph in \textbf{Figure \ref{fig:fig4}a}).
Increasing the temperature to 250 \degree C led to diffusion of solute species, as evidenced by the contrast changes in the STEM-HAADF micrograph in \textbf{Figure \ref{fig:fig4}b} and segregation of Ni at these features.
At 300 \degree C, the microstructure abruptly changes due to the formation of intermetallics (\textbf{Figure \ref{fig:fig4}c}), with isothermal annealing resulting in the dissolution of smaller intermetallics (\textbf{Figure \ref{fig:fig4}d}) through a ripening mechanism.
Curiously, the intermetallics formed in Al-2.3Ni-2.3Y following an \textit{in situ} annealing treatment differed from the intermetallics formed during the \textit{ex situ} annealing treatment, as in \textbf{Figure \ref{fig:fig3}c}, and more closely resemble the elongated complex intermetallics formed in the Ni-lean sample in \textbf{Figure \ref{fig:fig3}d}.
Such a behavior is a likely reflection of Ni diffusion to the lamella surfaces, resulting in Ni segregation observed in \textbf{Figure \ref{fig:fig4}b} and reducing the local Ni concentration, thereby promoting the formation of Ni-lean intermetallics at a lower retention temperature, 300 \degree C, reminiscent of the Ni-lean Al-1.3Ni-2.4Y. 
Thus, the instant formation of intermetallics at 300 \degree C, correlated with the exact temperature range at which the hardness drops, illustrates the direct connection between intermetallic formation and the shift in mechanical hardness. 
\par

To summarize, a combinatorial approach was utilized to examine the influence of the RE dopant in a ternary nanocrystalline Al-TM-RE alloy.
High-throughput characterization techniques, namely nanoindentation and XRD, were used to down-select ideal systems for probing microstructural transitions, and revealed two general regimes: an anneal-hardening and an anneal-softening regime.
In the anneal-hardening regime, nanocrystalline grain boundaries were stabilized through structural relaxation and dopant co-segregation towards the grain boundaries, increasing the overall hardness.
In the second regime, high temperatures drove the formation of intermetallics, creating possible pathways for strain localization and subsequent softening.
Differences in composition gave rise to variations in the retention temperature, as increasing the TM/RE ratio led to the formation of two intermetallic compositions with reduced distributions, facilitating softening through strain localization. 
In the Al-Ni-Y system, our results point to optimal thermal and mechanical stability when the Ni/Y ratio is close to 1, even when the overall Y content is as low as 2 at.\%, which is favorable when costs and sustainability considerations are important \cite{raabe2019strategies}.
Our results demonstrate the benefits of co-segregation of dopant species in ternary alloys, extending microstructural stability to higher temperatures while enhancing the mechanical hardness and encouraging the utilization of co-segregated compositions as the optimal compositions for microstructural stability.
\par


\section{Acknowledgements}
This research was sponsored by the Army Research Office under Grant Number W911NF-21- 1-0288. The views and conclusions contained in this document are those of the authors and should not be interpreted as representing the official policies, either expressed or implied, of the Army Research Office or the U.S. Government. The U.S. Government is authorized to reproduce and distribute reprints for Government purposes notwithstanding any copyright notation herein.  Support of the initial synthesis efforts is based upon work supported by the U.S. Department of Energy's Office of Energy Efficiency and Renewable Energy (EERE) under the Advanced Manufacturing Office Award Number DE‐EE0009114.



\bibliographystyle{elsarticle-num} 
\bibliography{refs}

\begin{thebibliography}{10}
\expandafter\ifx\csname url\endcsname\relax
  \def\url#1{\texttt{#1}}\fi
\expandafter\ifx\csname urlprefix\endcsname\relax\def\urlprefix{URL }\fi
\expandafter\ifx\csname href\endcsname\relax
  \def\href#1#2{#2} \def\path#1{#1}\fi

\bibitem{KUMAR:2003}
K.~Kumar, H.~{Van Swygenhoven}, S.~Suresh, Mechanical behavior of
  nanocrystalline metals and alloys, Acta Mater. 51~(19) (2003) 5743--5774.

\bibitem{Hahn:2015}
E.~N. Hahn, M.~A. Meyers, Grain-size dependent mechanical behavior of
  nanocrystalline metals, Materials Science and Engineering: A 646 (2015)
  101--134.

\bibitem{Padilla:2010}
H.~A. Padilla, B.~L. Boyce, A review of fatigue behavior in nanocrystalline
  metals, Exp. Mech. 50~(1) (2010) 5--23.

\bibitem{Rupert:2013}
T.~J. Rupert, W.~Cai, C.~A. Schuh, Abrasive wear response of nanocrystalline
  ni–w alloys across the hall–petch breakdown, Wear 298-299 (2013)
  120--126.

\bibitem{Beyerlein:2013}
I.~J. Beyerlein, A.~Caro, M.~J. Demkowicz, N.~A. Mara, A.~Misra, B.~P.
  Uberuaga, Radiation damage tolerant nanomaterials, Mater. Today 16~(11)
  (2013) 443--449.

\bibitem{ANDRIEVSKI:2014}
R.~Andrievski, Review of thermal stability of nanomaterials, Journal of
  materials science 49~(4) (2014) 1449--1460.

\bibitem{Chookajorn:2012}
T.~Chookajorn, H.~A. Murdoch, C.~A. Schuh, Design of stable nanocrystalline
  alloys, Science 337~(6097) (2012) 951--954.

\bibitem{Darling:2013}
K.~Darling, A.~Roberts, Y.~Mishin, S.~Mathaudhu, L.~Kecskes, Grain size
  stabilization of nanocrystalline copper at high temperatures by alloying with
  tantalum, Journal of Alloys and Compounds 573 (2013) 142--150.

\bibitem{Detor:2007}
A.~Detor, C.~Schuh, Microstructural evolution during the heat treatment of
  nanocrystalline alloys, Journal of Materials Research 22~(11) (2007)
  3233--3248.

\bibitem{El-Sherik:1991}
A.~El-Sherik, K.~Boylan, U.~Erb, G.~Palumbo, K.~Aust, Grain growth behaviour of
  nanocrystalline nickel, MRS Online Proceedings Library 238~(1) (1991)
  727--732.

\bibitem{Morris:1991}
D.~G. Morris, M.~Morris, Microstructure and strength of nanocrystalline copper
  alloy prepared by mechanical alloying, Acta metallurgica et materialia 39~(8)
  (1991) 1763--1770.

\bibitem{Cunningham:2022}
W.~S. Cunningham, S.~T. Mascarenhas, J.~S. Riano, W.~Wang, S.~Hwang, K.~Hattar,
  A.~M. Hodge, J.~R. Trelewicz, Unraveling thermodynamic and kinetic
  contributions to the stability of doped nanocrystalline alloys using
  nanometallic multilayers, Advanced Materials 34~(27) (2022) 2200354.

\bibitem{Amram:2018}
D.~Amram, C.~A. Schuh, Interplay between thermodynamic and kinetic
  stabilization mechanisms in nanocrystalline fe-mg alloys, Acta Materialia 144
  (2018) 447--458.

\bibitem{Chookajorn:2014}
T.~Chookajorn, C.~A. Schuh, Nanoscale segregation behavior and high-temperature
  stability of nanocrystalline w--20 at.\% ti, Acta materialia 73 (2014)
  128--138.

\bibitem{Khalajhedayati:2015}
A.~Khalajhedayati, T.~J. Rupert, High-temperature stability and grain boundary
  complexion formation in a nanocrystalline cu-zr alloy, Jom 67 (2015)
  2788--2801.

\bibitem{Koch:2008}
C.~Koch, R.~Scattergood, K.~Darling, J.~Semones, Stabilization of
  nanocrystalline grain sizes by solute additions, Journal of Materials Science
  43 (2008) 7264--7272.

\bibitem{XING:2018}
W.~Xing, A.~R. Kalidindi, D.~Amram, C.~A. Schuh, Solute interaction effects on
  grain boundary segregation in ternary alloys, Acta Materialia 161 (2018)
  285--294.

\bibitem{KUBE:2020}
S.~A. Kube, W.~Xing, A.~Kalidindi, S.~Sohn, A.~Datye, D.~Amram, C.~A. Schuh,
  J.~Schroers, Combinatorial study of thermal stability in ternary
  nanocrystalline alloys, Acta Materialia 188 (2020) 40--48.

\bibitem{LEI:2021}
T.~Lei, J.~Shin, D.~S. Gianola, T.~J. Rupert, Bulk nanocrystalline al alloys
  with hierarchical reinforcement structures via grain boundary segregation and
  complexion formation, Acta Materialia 221 (2021) 117394.

\bibitem{cantwell2014grain}
P.~R. Cantwell, M.~Tang, S.~J. Dillon, J.~Luo, G.~S. Rohrer, M.~P. Harmer,
  Grain boundary complexions, Acta Materialia 62 (2014) 1--48.

\bibitem{cantwell2020grain}
P.~R. Cantwell, T.~Frolov, T.~J. Rupert, A.~R. Krause, C.~J. Marvel, G.~S.
  Rohrer, J.~M. Rickman, M.~P. Harmer, Grain boundary complexion transitions,
  Annual Review of Materials Research 50 (2020) 465--492.

\bibitem{Balbus:2021}
G.~H. Balbus, J.~Kappacher, D.~J. Sprouster, F.~Wang, J.~Shin, Y.~M. Eggeler,
  T.~J. Rupert, J.~R. Trelewicz, D.~Kiener, V.~Maier-Kiener, D.~S. Gianola,
  Disordered interfaces enable high temperature thermal stability and strength
  in a nanocrystalline aluminum alloy, Acta Materialia 215 (2021) 116973.

\bibitem{Shin:2022}
J.~Shin, F.~Wang, G.~H. Balbus, T.~Lei, T.~J. Rupert, D.~S. Gianola, Optimizing
  thermal stability and mechanical behavior in segregation-engineered
  nanocrystalline al--ni--ce alloys: A combinatorial study, Journal of
  Materials Research 37~(18) (2022) 3083--3098.

\bibitem{Li:2018}
Z.~Li, A.~Ludwig, A.~Savan, H.~Springer, D.~Raabe, Combinatorial metallurgical
  synthesis and processing of high-entropy alloys, Journal of Materials
  Research 33~(19) (2018) 3156--3169.

\bibitem{Ding:2012}
S.~Ding, J.~Gregoire, J.~J. Vlassak, J.~Schroers, Solidification of au-cu-si
  alloys investigated by a combinatorial approach, Journal of Applied Physics
  111~(11) (2012) 114901.

\bibitem{li2021achieving}
Q.~Li, J.~Wang, H.~Wang, X.~Zhang, Achieving strong and stable nanocrystalline
  al alloys through compositional design, Journal of Materials Research (2021)
  1--25.

\bibitem{Balbus:2020}
G.~H. Balbus, F.~Wang, D.~S. Gianola, Suppression of shear localization in
  nanocrystalline al--ni--ce via segregation engineering, Acta Materialia 188
  (2020) 63--78.

\bibitem{gaultois2013data}
M.~W. Gaultois, T.~D. Sparks, C.~K. Borg, R.~Seshadri, W.~D. Bonificio, D.~R.
  Clarke, Data-driven review of thermoelectric materials: performance and
  resource considerations, Chemistry of Materials 25~(15) (2013) 2911--2920.

\bibitem{mansouri2017balancing}
A.~Mansouri~Tehrani, L.~Ghadbeigi, J.~Brgoch, T.~D. Sparks, Balancing
  mechanical properties and sustainability in the search for superhard
  materials, Integrating materials and manufacturing innovation 6 (2017) 1--8.

\bibitem{Cheary:hw0001}
R.~W. Cheary, A.~Coelho, {A fundamental parameters approach to X-ray
  line-profile fitting}, Journal of Applied Crystallography 25~(2) (1992)
  109--121.

\bibitem{Patterson:1939}
A.~L. Patterson, The scherrer formula for x-ray particle size determination,
  Phys. Rev. 56 (1939) 978--982.

\bibitem{Pun:2017}
S.~C. Pun, W.~Wang, A.~Khalajhedayati, J.~D. Schuler, J.~R. Trelewicz, T.~J.
  Rupert, Nanocrystalline al-mg with extreme strength due to grain boundary
  doping, Materials Science and Engineering: A 696 (2017) 400--406.

\bibitem{Davis:1993}
J.~R. Davis, Aluminum and aluminum alloys, ASM international, 1993.

\bibitem{mantha2021grain}
L.~S. Mantha, B.~E. MacDonald, X.~Mu, A.~Mazilkin, J.~Ivanisenko, H.~Hahn,
  E.~J. Lavernia, S.~Katnagallu, C.~K{\"u}bel, Grain boundary segregation
  induced precipitation in a non equiatomic nanocrystalline cocufemnni
  compositionally complex alloy, Acta Materialia 220 (2021) 117281.

\bibitem{devaraj2019grain}
A.~Devaraj, W.~Wang, R.~Vemuri, L.~Kovarik, X.~Jiang, M.~Bowden, J.~Trelewicz,
  S.~Mathaudhu, A.~Rohatgi, Grain boundary segregation and intermetallic
  precipitation in coarsening resistant nanocrystalline aluminum alloys, Acta
  Materialia 165 (2019) 698--708.

\bibitem{zhao2018segregation}
H.~Zhao, F.~De~Geuser, A.~K. da~Silva, A.~Szczepaniak, B.~Gault, D.~Ponge,
  D.~Raabe, Segregation assisted grain boundary precipitation in a model
  al-zn-mg-cu alloy, Acta Materialia 156 (2018) 318--329.

\bibitem{VASILIEV:2004}
A.~Vasiliev, M.~Aindow, M.~Blackburn, T.~Watson, Phase stability and
  microstructure in devitrified al-rich al–y–ni alloys, Intermetallics
  12~(4) (2004) 349--362.

\bibitem{Mika:2010}
T.~Mika, B.~Kotur, Phase equlibria in the {Y, Gd}–ni–al ternary systems in
  the 65-100 at.\% al range at 773 k: a reinvestigation, Chemistry of Metals
  and Alloys 3 (2010).

\bibitem{Cordero:2016}
Z.~C. Cordero, B.~E. Knight, C.~A. Schuh, Six decades of the hall–petch
  effect – a survey of grain-size strengthening studies on pure metals,
  International Materials Reviews 61~(8) (2016) 495--512.

\bibitem{raabe2019strategies}
D.~Raabe, C.~C. Tasan, E.~A. Olivetti, Strategies for improving the
  sustainability of structural metals, Nature 575~(7781) (2019) 64--74.

\end{thebibliography}

\end{document}